\documentclass[12pt]{article}

\usepackage[utf8]{inputenc}

\usepackage[english, russian]{babel}
\usepackage{ytableau}
\usepackage{braket}
\usepackage{amsmath}
\usepackage{mathtools}
\usepackage{breqn}
\usepackage{graphicx}
\usepackage{amsfonts}
\usepackage{indentfirst}
\usepackage{amssymb}

\makeatletter \@addtoreset{equation}{section} \makeatother

\renewcommand{\d}{\mbox{d}}
\newcommand{\un}{{\underline{n} }}
\begin{document}

\begin{flushright}
%\texttt{Started on May 14}
%\\
{\small FIAN/TD/16/2022}
\end{flushright}
\vspace{1.7 cm}

\begin{center}
{\large\bf Unfolded Point Particle as a Field in Minkowski Space}

\vspace{1 cm}

{\bf A.A. Tarusov${}^{1,2}$ and   M.A.~Vasiliev${}^{1,2}$}\\
\vspace{0.5 cm}
{\it
${}^1$ I.E. Tamm Department of Theoretical Physics, Lebedev Physical Institute,\\
Leninsky prospect 53, 119991, Moscow, Russia}

\vspace{0.7 cm} \textit{ ${}^2$
Moscow Institute of Physics and Technology, Institutsky pereulok 9, 141701, Dolgoprudny, Moscow region, Russia}
\vspace{0.6 cm}
% vasiliev@lpi.ru \\
\end{center}

	\selectlanguage{english}
\renewcommand{\abstractname}{Abstract}
\begin{abstract}

Point-particle dynamics is reformulated as a field theory. This is
achieved by using the unfolded dynamics approach that makes it possible
to give dynamical interpretation to the concept of physical dimension
which is 1 for a point particle in the $d$-dimensional space-time.
The main idea for the description of a $k$-dimensional on-shell system
in the $d$-dimensional space is to keep the evolution along $d-k$ dimensions
off-shell or, alternatively, restrict it in a specific way
respecting the compatibility conditions of the resulting unfolded system.
The developed approach gives some hints how a non-linear realization of the
symmetry $G$ of a larger-dimensional space in a lower-dimensional system can
emerge from a geometrical realization on the fields in an appropriate $G$-invariant
space. For the example of a relativistic point particle considered in this paper,
$ G$ is the Poincar\'e group. The proposed general scheme is illustrated by simple
examples that reproduce conventional results.

\end{abstract}
\newpage
\renewcommand{\contentsname}{Table of contents}
\tableofcontents

\newpage
\section{Introduction}
Among different approaches to relativistic theories, one can distinguish between the
world-line particle approach and the field-theoretic one. In this article, we unify  these approaches within the unfolded formulation of dynamical systems \cite{Vasiliev:1988xc,Vasiliev:1988sa}, which will allow us to ascribe a dynamical sense to physical dimensions of a system \cite{Vasiliev:2001dc,Vasiliev:2014vwa}.

An example  of this phenomenon was given in \cite{Vasiliev:2001zy}, where, following Fronsdal's prescription \cite{F}, an infinite system of massless fields of all spins in four-dimensional space has been described by a single field in the ten-dimensional space (Analogous results were achieved in the particle approach somewhat earlier in \cite{BLS}).

In this approach  space-time geometry in which the dynamical equations are formulated is determined by the symmetry acting on the space, while the physical dimension is associated with the set of initial data, that determine the evolution of the system \cite{Vasiliev:2001dc}. In the papers \cite{Vasiliev:2001zy,Vasiliev:2001dc,BLS} (see also \cite{Bandos:2005mb}) this idea was realized for the symmetry acting linearly on the fields in both the four- and ten-dimensional space.

The description of the dynamics of point particles elaborated in this article assumes a non-linear realization of the Lorentz symmetry on the dynamic variables of the point particle as a consequence of the Einstein constraint on the  velocity four-vector
$$
u^n u_n = 1\,.
$$
(By selecting any $u^n$, that obeys this constraint, the Lorentz symmetry gets
spontaneously broken.)
The application of the unfolded formalism in this case differs significantly from the linear case.
It requires the introduction of additional fields which encode unconstrained or specific evolution along
"extra" dimensions of space-time as is explained in this paper.

The paper is organised as follows:
Section \ref{Unfolding} recalls the unfolded formalism used in the paper, illustrated by the well-known
scalar field example. Section \ref{FreeParticle} details the application of that formalism to the case of a free point particle. In Section \ref{Off-shell system} an off-shell formulation of the unfolded particle dynamics
 is presented both in terms of component fields and in terms of generating functions. In Section \ref{onshell}
 it is explained how  an external force to the equations of motion can be  introduced and a number of simple examples of the on-shell systems is considered.  Section \ref{Conclusions} contains brief conclusions
 with some emphasize on the further applications and open problems.
%\newpage
\section{Unfolding} \label{Unfolding}
The possibility of decreasing the order of a differential equation by introducing new variables and transitioning to an equivalent system of differential equations is well known.
Extension of this approach to partial differential equations is based on the jet formalism \cite{Vinogradov}.
Also it was elaborated in the framework of  BV-BRST formalism  e.g. in \cite{Batalin:1977pb,Barnich:2004cr,Barnich:2010sw,Grigoriev:2012xg}.

Unfolded dynamics approach \cite{Vasiliev:1988xc,Vasiliev:1988sa} (see also \cite{Boulanger:2008up}), that is most appropriate for the gauge theories in the framework of gravity, is a generalization of the  first-order
formulation of a system  via replacing a partial derivative by de Rham derivative $\d := \xi^\un
\frac{\partial}{\partial x^\un}$
and dynamical variables by space-time differential forms $W(x)$, which allows one to rewrite the system of equations in the form
\begin{equation}
\label{dw}
\d W^\Omega(\xi^\un, x) = G^\Omega(W(\xi^\un, x))\,,
\end{equation}
where  $\xi^\un$ is the anticommuting  differential used as a placeholder for $dx^\un$, and $G^\Omega(W(\xi, x))$ is  some function of
$W$ containing only exterior products of the differential forms $W(\xi, x)$ at the same $x$
(no space-time derivatives in $G^\Omega(W(\xi, x))$; wedge products are implicit). The functions $G^\Omega(W(\xi, x))$ cannot be arbitrary,
as the de Rham derivative is nilpotent and thus the compatibility condition $\d G^\Omega=0$ must hold. This demands
\begin{equation}
\label{con}
G^\Lambda( W) \frac{\partial G^\Omega(W)}{\partial W^\Lambda} = 0\,.
\end{equation}
This constraint allows one to show that
 system (\ref{dw}) is manifestly invariant under
 the following gauge transformations:
\begin{equation}
	\delta_{gauge}  W^\Omega(\xi^\un, x) = \d \epsilon^\Omega(\xi^\un, x) + \epsilon^\Lambda(\xi^\un, x) \frac{\partial G^\Omega(W(\xi^\un,x))}{\partial W^\Lambda (\xi^\un, x)}\,, \label{gaugetransform}
\end{equation}
where
$$
{\rm deg}\,\epsilon^\Lambda(\xi^\un, x) = {\rm deg}\, W^\Lambda (\xi^\un, x) - 1 \,,
$$
with ${\rm deg}\, \omega$ being a differential form degree of $\omega$.

Generally speaking, gauge invariance only takes place in so-called universal systems
\cite{Bekaert:2005vh,Vasiliev:2005zu}, in which the compatibility conditions hold as a consequence of the system itself without taking into account the number of space-time dimensions, i.e.  the fact that $d+1$-forms vanish in $d$-dimensional space. Indeed, for non-universal systems  partial derivative $\frac{\partial G^\Omega}{\partial W^\Lambda}$ might  have no sense, leading to a non-zero derivative of zero represented by a $d+1$-form.

In other terms, the fact that $G^\Lambda(\xi^\un, x)$ is a function of $W^\Omega(\xi^\un, x)$  can be written as
\begin{equation}
	 G^\Lambda = \sum_{n=1}^{\infty} f^\Lambda{}_{\Omega_1, ... \Omega_n} W^{\Omega_1} ... W^{\Omega_n}  \,,\qquad
	 f^\Lambda{}_{\Omega_1, ... \Omega_k , \Omega_{k+1}, ... \Omega_n} = (-1)^{deg_{\Omega_k} deg_{\Omega_{k+1}}} f^\Lambda{}_{\Omega_1, ... \Omega_{k+1}, \Omega_k, ... \Omega_n} \,.
\end{equation}
(From now on we  omit arguments of $G^\Lambda$ and $W^\Omega$ if it does not lead to misunderstandings.)
Compatibility condition (\ref{con}) then yields generalized Jacobi identities
on the structure constants $f$
\begin{equation}
	\sum_{n=0}^{m} (n+1) f^\Lambda{}_{[\Omega_1, ... \Omega_{m-n}} f^\Phi{}_{\Lambda, \Omega_{m-n+1}, ... \Omega_{m}  \}} = 0 \,,
\end{equation}
where $[\}$ indicates an  appropriate (anti)symmetrisation of indices. From this point of view, the universality of a system means that the generalized Jacobi identities hold true regardless of the dimension $d$. The underlying mathematical structure is called the strong homotopy $L_\infty$ algebra  \cite{Lada:1992wc}.
%The systems considered in this paper are indeed universal.

For universal systems it is also possible to introduce a $Q$-differential
(homological vector field) of the  form \cite{Vasiliev:2005zu}
\begin{equation}
Q = G^\Omega \frac{\partial}{\partial W^\Omega}\,,
\end{equation}
which turns out to be nilpotent,
$$
Q^2 =0\,,
$$
as a consequence of compatibility conditions (\ref{con}). In these terms, any universal unfolded system can be rewritten as
$$\d F(W)= QF(W)\,,$$
where $F(W)$ is an arbitrary function.
This way of describing the system as a so-called $Q$-manifold relates the de Rham derivation on the world-sheet with coordinates $x^\un$ to the derivation $Q$ on the target space with coordinates $W^\Lambda$.

Unfolded formalism allows for a natural way of description of background geometry via Maurer–Cartan equations which have the unfolded form. Indeed, let $g$ be a Lie algebra. Setting $W = w$ and $G=  \frac{1}{2} [w\,,w]$ for a one-form $w\in g$, one observes that equation (\ref{dw}) yields the zero-curvature condition
\begin{equation}
\d w + \frac{1}{2} [w\,,w]= 0\,. \label{flatcond}
\end{equation}
For $g$ being  Poincar\'e algebra with the one-form gauge fields (connection) $e^n$ and $w^{nm}$, the unfolded system
(\ref{flatcond}) yields the coordinate-independent description of  Minkowski space in the form
\begin{equation}
	\mathcal{R}^n= 0 \,,\qquad
	\mathcal{R}^{nm}= 0 \,.
\end{equation}

Consider a system of equations $$\mathcal{D} C_A (x) = 0\,,$$
where the fields $C_A (x)$ in Minkowski space are valued in a Poincar\'e-module $V$ while $\mathcal{D}$ is the
exteriour covariant
derivative in $V$
with the flat connection $w = e^n P_n + \omega^{nm}L_{nm}$ obeying (\ref{flatcond}): $\mathcal{D} = d + w$. (For instance, Lorentz transformations
 act on the tensor indices.) This system is
 invariant under the following gauge transformations:
\begin{align}
& \delta C_A = -\varepsilon_A{}^B C_B, \\
& \delta w (x) = \mathcal{D} \varepsilon (x).
\end{align}
From here it follows that for a fixed $w = w_0$ the gauge symmetry parameters are restricted by the condition
\begin{equation}
 \mathcal{D}_0 \varepsilon (x) = 0 \label{wtransf}\,,\qquad
 \mathcal{D}_0 = \d + w_0.
\end{equation}
Since $\mathcal{D}_0^2 = 0$, in the topologically trivial case the zero-form parameter $\varepsilon_B{}^A$, can be reconstructed from any point hence generating global symmetries of the system.

Minkowski space in Cartesian coordinates is described by the connection
\begin{equation}
  e^n=dx^n\,,\qquad\omega^{nm}=0\,,
\end{equation}
in which case (\ref{wtransf}) yields
\begin{equation}
	\partial_n \varepsilon^n - \varepsilon^n{}_m e_n{}^m = 0 \,, \qquad
	\partial_n \varepsilon^{nm} = 0 \label{poincare2}\,,
\end{equation}
which can be solved as $\varepsilon^{nm} = - \varepsilon^{mn} = const$, $\varepsilon^n = \varepsilon^n{}_m x^m + \epsilon^n$, where $\epsilon^n$ is $x$-independent. This  obviously forms the Poincar\'e transformations.

%\newpage
\section{Free particle} \label{FreeParticle}
The classical point particle is conventionally  described by generalized coordinates $q^i(s)$ depending on the evolution parameter $s$. In this paper, we also use the generalized coordinates $q^i = q^i(x)$ which, however, will be treated as space-time fields, i.e., functions of all
space-time coordinates $x^n$. Let us, for simplicity, work with  Cartesian coordinates $x^n$ of a flat Minkowski space, which will allow us to omit the Lorentz connection dependent terms.

We start with a first-step equation
\begin{equation}
Dq^i(x) = e^j q^i{}_j (x)\,,
\end{equation}
where  $q^i{}_j(x)$  is an arbitrary (for now) matrix and $D$ is the Lorentz covariant derivative. In
Cartesian coordinates this yields
\begin{equation}
\d q^i (x) = e^j q^i{}_j (x). \label{freeq}
\end{equation}

At $j=0$  one arrives at a differential equation with respect to time $t=x^0$ with an arbitrary right hand side.
However, in contrast with classical dynamics, other values of $j$ also produce non-trivial equations, which means that the equations of motion involve all space-time variables.
This unusual modification makes sense when treating generalized coordinates as embedding functions from our laboratory system to some other. In this case the space sector of the right hand side is just a Jacobian of that transformation. It is convenient to assume that the dimensions of the original and target spaces are the same, thus the space sector of our matrix has to be non-degenerate.
Apart from the dependence on "extra" variables, this is  similar to the description of a classical particle in terms of the transformation from a chosen reference frame  to the one in which the particle is at rest.

The analysis does not stop here though, since  $q^i{}_j$ must satisfy the compatibility condition
\begin{equation}
\label{compa}
e^j Dq^i{}_j (x) = 0\,.
\end{equation}
The general solution to this condition has the form of a one-form with tensor coefficients that are symmetric in lower indices, which solves (\ref{compa}) due to anticommutativity of the one-forms $e^n$, $e^i e^j = - e^j e^i$,
\begin{equation}
\label{1}
 Dq^i{}_j (x) = e^k q^i{}_{jk} (x)\,,\qquad
 q^i{}_{jk} (x) = q^i{}_{kj} (x)\,.
\end{equation}

Equation (\ref{1}) also produces the compatibility condition which has the analogous form
\begin{equation}
Dq^i{}_{jk} (x) = e^l q^i{}_{(jkl)} (x)\,.
\end{equation}
The process can be continued resulting in the infinite set of equations
\begin{align}\label{unfq}
& Dq^i{}_{(j_1 ... j_n)} (x) = e^k q^i{}_{(j_1 ... j_n k)} (x)\,.
\end{align}

This system provides an example of an off-shell  unfolded
system that does not describe any non-trivial equations of motion. It is fully analogous to
that described in \cite{Shaynkman:2000ts} for the scalar field case. (Supersymmetric extensions
of the off-shell unfolded systems were recently considered in \cite{Misuna:2022zjr}.)
Every equation
expresses the compatibility of the previous one, but involves a new object that requires its own compatibility condition.
To get nontrivial dynamics, however, one has to introduce additional
conditions on the coefficients $q^i{}_{(j_1 ... j_n k)}$ describing
higher derivatives of the field $q^i$.
Imposing constraints on the fields $q^i{}_{(j_1 ... j_n k)}$ is equivalent to
imposing some differential equations on the fields $q^i$.

The equations (\ref{unfq})  admit a more compact form using auxiliary variables $y^i$, so that the right hand side of the equations results from differentiation of the generating functions with respect to $y^i$,
\begin{equation}
	Dq^i{}_{(j_1 ... j_n)} (x, y) = e^k \frac{d}{dy^k} q^i{}_{(j_1 ... j_n)} (x, y)\,, \label{GeneratingFunctionOrigin}
\end{equation}
where $q^i(x,y)$ is the generating function
\begin{align}
	q^i(x,y) = \sum_{n=0}^{\infty} \frac{1}{n!} q^i{}_{j_1, ... j_n} (x) y^{j_1} ... y^{j_n} \,,
\end{align}
with the original field $q^i(x)$  recovered at $y = 0$.

To describe a free relativistic particle this way we introduce a time-like ``velocity 4-vector" $V^i(x)$
as a new variable, imposing the equations
\begin{align}
& Dq^i (x) = e^j V_j (x) V^i (x), \label{dqevv}\\
& DV^i (x) = 0, \label{DV0}\\
& V^i (x) V_i (x) = 1. \label{absV}
\end{align}

Let us show that this system  indeed describes a free relativistic particle.
In Cartesian coordinates the system takes the form
\begin{align}
& e^j \frac{\partial}{\partial x^j} q^i (x) = e^j V_j (x) V^i (x),\\
& e^j \frac{\partial}{\partial x^j} V^i (x) = 0.
\end{align}
Here the second equation implies that $V^i$ is a  constant
while the first one contains a one-form
\begin{equation}
\kappa = e^i V_i \label{DefineKappa}\,,
\end{equation}
which, in a sense, serves as a projector on the world line of the particle.

There is some freedom in the parametrization of  the world line. For example, to take time $x^0$ as the
evolution parameter, one has to reduce  $dx^n$ to $dx^0$ (equivalently,
$e^n \rightarrow dx^{\underline{n}} \delta^0_{\underline{n}}$). This reproduces the familiar equations of motion.
Indeed, after such a reduction, the velocity vector produces a factor of $V_0$, which is just a relativistic gamma-factor $\gamma = (1- \frac{v^2}{c^2})^{-1 / 2}$. This is not surprising, since the differentiation on the left hand side is over laboratory time,
\begin{align}
& \dot{q}^i (x) = \gamma V^i (x), \label{Reductionq}\\
& \dot {V}^i (x) = 0. \label{Reductionv}
\end{align}
Thus, equations (\ref{dqevv})-(\ref{absV}) indeed describe propagation of a free point particle with the 4-velocity $V^i(x)$.

Since our unfolded system can be easily extended  to include the Lorentz connection
by appending (\ref{flatcond}), it inherits the full Poincar\'e symmetries
as outlined at the end of Section 2.
The unfolded formalism allows us to straightforwardly derive the symmetries of the system. In this case (\ref{gaugetransform}) generates the background Poincar\'e transformation as well as a
transformation of $q^i$
\begin{equation}
	\delta q^i = \prescript{0}{}{\epsilon}^j  \frac{\partial e^k V_k V^i}{\partial e^j} = \prescript{0}{}{\epsilon}^j V_j V^i \,.
\end{equation}

Note that in this formalism nontrivial particle dynamics is only along the direction associated with
 $\kappa$. In other (transversal) directions the dynamics is trivial with no dependence
on the other coordinates.

%\newpage
\section{Off-shell system}
\label{Off-shell system}
\subsection{General setup}
\label{General setup}
The world-line  one-form $\kappa$ (\ref{DefineKappa}) makes it possible to formulate
the off-shell unfolded system of a specific form distinguishing between the directions along $\kappa$ and transversal ones. The evolution of the system in the transversal directions is necessary for
 consistency. Indeed, the naive system
\begin{equation}
%&
 Dq^i (x) = \kappa V^i (x)\label{naivesystem1}\,,\qquad
%\\&
DV^i (x) = \kappa F^i (x)\,, %\label{naivesystem2}
\end{equation}
is inconsistent for arbitrary $F^i$ because now
 $\kappa$ is not closed
\begin{equation}
D\kappa = - e^i D V_i = - e^i \kappa F_{i} \,.
\end{equation}

While the classical behavior of the system is defined by the terms aligned with $V^i$, to achieve compatibility  in all directions one has to adjust the evolution along the transversal directions appropriately.

To achieve this it is convenient to introduce the transversal one-forms
\begin{align}
\eta^i := e^i - \frac{\kappa V^i}{V^2} \,,\qquad V^i \eta_i =0\,,
\end{align}
where an additional normalization is introduced, since the condition (\ref{absV}) is relaxed, as it is not necessarily true off-shell. (Still we assume that $V^2(x):= V^i(x)V_i(x)\neq 0$.) The system then takes the
form
\begin{align}
& Dq^i = \kappa V^i (x) + \eta^j H^i{}_j(x), \label{DqWithPerp}\\
& DV^i = \kappa F^i (x)  + \eta^j G^i{}_j(x). \label{DVWithPerp}
\end{align}

Since $\eta^i$ is $V^i$-transversal, this system is invariant under the
``gauge" transformations
\begin{align}
	& H^{'i}{}_j(x) = H^i{}_j(x) + \phi^i(x) V_j(x), \\
	& G^{'i}{}_j(x)= G^i{}_j(x)  + \psi^i(x) V_j(x)\,,
\end{align}
with arbitrary functions $\phi^i (x)$, $\psi^i (x)$, that can be gauge fixed by demanding
\begin{align}
	& H^i{}_j V^j = 0\,, \qquad
	G^{i}{}_j V^j = 0 \,.
\end{align}

Firstly, let us note that the system is indeed off-shell as long as the condition $V^i V_i=1$ is not enforced. Indeed, the left hand sides contain $d^2$ components of first derivatives of $q^i(x)$  (or $V^i(x)$ for the second equation). On the right hand side, the $H^i{}_j$  ($G^i{}_j$) contain  $d(d-1)$ components due to the transversality condition while the $V^i$  ($F^i$) span the $d$ leftover  components.

To check the compatibility conditions of this system, one has to act by $D$ on the both sides of the equations then solving them with respect to $G^i{}_j$ and $H^i{}_j$. We analyze the system in an arbitrary torsion free
geometry with  $De^i=0$, which yields
\begin{align}
	& D\kappa = - e^i \kappa F_{i}  -  e^i \eta^j G_{ij}, \\
	& D \eta^i = \frac{1}{V^4} \left((e^k \kappa F_{k} V^i + e^k \eta^j G_{kj} V^i + \kappa \eta^j G^i{}_j) V^2 - 2 \kappa V^i V_k \eta^j G^k{}_j \right).
\end{align}

The compatibility of (\ref{DqWithPerp}) yields using $DD A^i=R^i{}_k A^k = e^l e^j R^i{}_{k,lj} A^k \,. $
\begin{multline}
	DDq^i = (D\kappa) V^i (x) - \kappa D V^i (x) + D(\eta^j) H^i{}_j(x) -  \eta^j D H^i{}_j(x) = \\ = - V^i e^j \kappa F_{j}  - V^i e^k \eta^j G_{kj} - \kappa \eta^j G^i{}_j +  \frac{1}{V^2} \kappa \eta^l G^j{}_l  H^i{}_j - \eta^j DH^i{}_j =  e^l e^j R^i{}_{k,lj} q^k\,.
\end{multline}
Expanding the last equation in the basis  two-forms $\kappa \eta^i$ and $\eta^i\eta^j$, we obtain
\begin{dmath}
	\eta^j DH^i{}_j = - V^i \eta^j \kappa F_{j} - \frac{1}{V^2} V^i \kappa V^k \eta^j G_{kj} - V^i \eta^k \eta^j G_{kj} - \kappa \eta^j G^i{}_j + \frac{1}{V^2}\kappa \eta^l G^j{}_l  H^i{}_j -  \frac{2}{V^2} \kappa V^l \eta^j R^i{}_{k,lj} q^k - \eta^l \eta^j R^i{}_{k,lj} q^k\,,
\end{dmath}
which is equivalent to
\begin{dmath}
	DH^i{}_j = \kappa (- F_{j} V^i + G^i{}_j +  \frac{1}{V^2} V^i V^k G_{kj} - \frac{1}{V^2} G^k{}_j H^i{}_k + \frac{2}{V^2} V^l R^i{}_{k,lj} q^k)  + \eta^k G_{kj} V^i + \eta^l R^i{}_{k,lj} q^k + \eta^k A^i{}_{jk} + \kappa V_j B^i + V_j \eta^k C^i{}_k, \label{DH}
\end{dmath}
where the last three terms  with arbitrary $B^i$, $C^i{}_k$ and symmetric $A^i{}_{jk} = A^i{}_{kj}$
parameterize the general solution of  the homogeneous equation $\eta^j DH^i{}_j = 0$. Just as for $H^i{}_j$ itself, the transversality condition can be imposed on  $A^i{}_{jk}$ and $C^i{}_k$,
\begin{align}
	A^i{}_{jk} V^k = 0	\,,\qquad
	C^i{}_k V^k = 0 \,.
\end{align}

Analogously for equation (\ref{DVWithPerp}), with the only difference that we now impose the unfolded equations on
the field $F^i$,
\begin{equation}
	DF^i = \kappa J^i + \eta^j K^i{}_j \,
\end{equation}
again demanding $K^i{}_j V^j = 0$. Then the compatibility condition for $V^i$ yields
\begin{multline}
	DDV^i = (D \kappa) F^i  (x) - \kappa (DF^i )  + (D\eta^j) G^i{}_j - \eta^j (DG^i{}_j) = \\ = - e^j \kappa F_{j} F^i  - e^l \eta^j G_{lj} F^i - \kappa \eta^j  K^i{}_j + \frac{1}{V^2} \kappa \eta^k G^j{}_k G^i{}_j- \eta^j DG^i{}_j = e^l e^j R^i{}_{k,lj} V^k \,,
\end{multline}
or, equivalently,
\begin{dmath}
	\eta^j DG^i{}_j = \eta^j \kappa (- F_{j} F^i + K^i_j + \frac{1}{V^2} F^i V^k G_{kj} - \frac{1}{V^2} G^k{}_j G^i{}_k + \frac{2}{V^2} V^l R^i{}_{k,lj} V^k) + \eta^j \eta^k (G_{kj} F^i + R^i{}_{l,kj} V^l).
\end{dmath}

The solution again consists of the inhomogeneous part and the terms parameterizing a general
solution of the homogeneous equation,
\begin{dmath}
	DG^i{}_j = \kappa (- F_{j} F^i + K^i_j + \frac{1}{V^2} F^i V^k G_{kj} - \frac{1}{V^2} G^k{}_j G^i{}_k + \frac{2}{V^2} V^l R^i{}_{k,lj} V^k) +  \eta^k (G_{kj} F^i + R^i{}_{l,kj} V^l) + \eta^l M^i{}_{jl} +  \kappa V_j N^i + V_j \eta^k L^i{}_k\, \label{DG}
\end{dmath}
with $M^i{}_{jl}=M^i{}_{lj}$. Once again, $M^i{}_{jl}$ and $L^i{}_k$ obey the transversality conditions
\begin{align}
	M^i{}_{jk} V^k = 0	\,,\qquad
	L^i{}_k V^k = 0 \,.
\end{align}

In its turn, consistency of equations (\ref{DH}) and (\ref{DG})  imposes differential
constraints on the yet unconstrained coefficients $A,B,C$ and $M,N,L$ in terms of new
unconstrained variables. This process continues
indefinitely  leading eventually to a totally consistent infinite set of equations on the infinite set of
variables. Since the analysis of all these conditions in terms of component fields like $H,G,A,B,C,M,N,L$
quickly gets complicated we now revisit them  in a more compact form of generating functions.

\subsection{Covariant constraints} \label{CovCon}
Though the description of a point particle considered in Section \ref{General setup} is
clear in principle it is algebraically involved and not instructive.
It  can be simplified at least in Minkowski background by imposing appropriate constraints
in terms  of generating functions of Section \ref{FreeParticle}.
To this end, we introduce auxiliary variables $y^i$ as  in (\ref{GeneratingFunctionOrigin}), rewriting the system (\ref{DqWithPerp}), (\ref{DVWithPerp}) as
\begin{align}
& Dq^i(x,y) = e^j \frac{d}{dy^j} q^i(x,y), \label{Dqy}\\
& DV^i(x,y) = e^j \frac{d}{dy^j} V^i(x,y). \label{DVy}
\end{align}
These equations are clearly consistent, as derivatives commute while the vielbein one-forms anticommute. They  do not describe any dynamics, imposing no conditions on $q^i(x,0)$ and $V^i(x,0)$. The results of
 Section \ref{General setup} can be reproduced by  imposing the following conditions:
\begin{equation}
V^i (x,y) \frac{d}{dy^i} q^j (x,y) = V^2(x,y) V^j(x,y) \label{Vdq0}\,.
\end{equation}

One has  to check that this constraint is compatible with  (\ref{Dqy}), (\ref{DVy}), i.e. its differentiation does not produce new constraints, giving zero by virtue of (\ref{Vdq0}). Indeed,
\begin{equation}
 D\Big (V^i (x,y) \frac{d}{dy^i} q^j (x,y) - V^2 (x,y)V^j(x,y)\Big) = e^k \frac{d}{dy^k} \Big (V^i (x,y) \frac{d}{dy^i} q^j (x,y) - V^2(x,y) V^j(x,y)\Big) = 0 ,
\end{equation}
(In the sequel, the arguments of the generating functions $q^i(x,y)$, $V^i(x,y)$ are implicit.)

Let us now show that supplemented with constraint (\ref{Vdq0}) equations (\ref{Dqy}), (\ref{DVy}) reproduce the equations from the previous section.
By virtue of (\ref{Vdq0}), and since $ e^i = \frac{\kappa V^i}{V^2} + \eta^i$, Eq.~(\ref{Dqy}) yields
\begin{equation}
Dq^i = \kappa V^i + \eta^j \frac{d}{dy^j}q^i \,.
\end{equation}
 Equation (\ref{DqWithPerp}) is reproduced
with $\eta^j \frac{d}{dy^j}q^i\vert_{y=0} = \eta^j H^i{}_j$. To fix an obvious freedom up to a function $\phi^i V_j$ in a way preserving trasversality one can set
$$ H^i{}_j=
\Big (\frac{d}{dy^j}q^i - \frac{1}{V^2} V_j V^k\frac{d}{dy^k}q^i\Big ) \Big \vert_{y=0} \,.$$

Analogously, (\ref{DVy}) yields equation (\ref{DVWithPerp}) with
\begin{align}
	& V^j \frac{d}{dy^j}V^i \Big \vert_{y=0} = F^i \,, \\
	&\Big ( \frac{d}{dy^j}V^i - \frac{1}{V^2} V_j V^k \frac{d}{dy^k}V^i \Big )\Big\vert_{y=0}= G^i{}_j\,.
\end{align}

Let us note that the second derivative of the generating function has $\frac{d^2(d+1)}{2}$ independent components, of which, keeping in mind the transversality conditions, $\frac{d^2(d+1)}{2} - d^2$ are encoded in $A^i{}_{jk}$, $d^2 -d$ in $C^i{}_k$ and $d$ more in $B^i$. That means that the system (\ref{Dqy})-(\ref{Vdq0})
indeed concisely reproduces the off-shell formulation of Section \ref{General setup} in all orders.

\section{Examples of on-shell systems}
\label{onshell}

To put the system on-shell one has to set the field $F^i$, that determines the evolution of $V^i$ along itself,  to some function $F^i(q, V)$. Restriction of some combination of derivatives parameterized by $F^i$
then would impose some partial differential equations on $q^i$ giving rise to the equations of motion.
Generally, a non-zero force $F^i(q, V)$ would demand some higher components of additional fields associated with the higher components in $y^j$ of $q^i(x,y)$ and $V^i(x,y)$ (descendants) to be nonzero. There are two somewhat opposite options.

One is that all these descendants are kept non-zero and arbitrary in the sense that they parameterize a general solution to the compatibility conditions. Another one is that these descendants
give as simple as possible specific solution to the compatibility conditions.
In the former case the system turns out to be off-shell in all directions transversal to the trajectory.
In the latter, the evolution along transversal directions has a specific form compatible with
$F^i(q, V)\neq 0$ in the full unfolded system.
Postponing a general analysis of this issue for the future publication here we consider a few simple
examples of the second kind.

\subsection{Lorentz force in a constant field} \label{LorentzForce}

A particular choice of $F^i (q, V)$ linear in $V$, $F^i (q,V) = F^i{}_j(q) V^j$, $F_{ij} = - F_{ji}$, replicates the Lorentz force. As a toy example consider a particular solution to the compatibility conditions of (\ref{DqWithPerp}), (\ref{DVWithPerp}) in flat space, that  easily puts the system on-shell. Namely, let
$F^i{}_j(q)$ be a constant field, {\it i.e.} $\d F^i{}_j=0$. Antisymmetry of $F_{ij}$ allows us to impose condition (\ref{absV})
and write down the following on-shell system:
\begin{align}
	& \d q^i(x) = \kappa V^i + \eta^i = e^i, \label{Lorentzq} \\
	& \d V^i(x) = \kappa V^j F^i{}_j + \eta^j F^i{}_j = e^j F^i{}_j, \label{LorentzV}
\end{align}
which is obviously consistent without introducing higher components in $y^j$ of
$q^i(x,y)$ and $V^i(x,y)$.

Note that the free particle case considered in Section \ref{FreeParticle}  is reproduced at $F^i=0$ and
 also corresponds to the
specific (trivial) choice of the descendants associated with higher components in $y^j$ of
$q^i(x,y)$ and $V^i(x,y)$.

\subsection{Gravitational interaction}

Within the exterior algebra formalism underlying the unfolded dynamics approach,
 the gravitational background is naturally taken into account by using appropriate covariant derivatives of the Cartan formulation of gravity. To introduce it in the metric formalism, i.e. with Christoffel symbols, one has to distinguish between laboratory Lorentz indices denoted by Latin letters and the underlined world sheet indices. For instance,  $V^i =e_{\protect\underline{i}}{}^i V^{\protect\underline{i}}$, where the vielbein $e_{\protect\underline{i}}{}^i$ relates laboratory and world indices. Let us start with the Cartan formulation.

The on-shell covariant condition (\ref{DVWithPerp}) for $V^i$ with zero force reads as
\begin{equation}
	\partial_{\protect\underline{k}}  V^i = - \omega_{\protect\underline{k}}{}^i{}_j V^j + \eta_{\protect\underline{k}}{}^j G^i{}_j\,. \label{SpinConnectionFormulation}
\end{equation}
On-shell, it is possible to impose the condition (\ref{absV}), compatibility with which then implies the anticipated antisymmetry of  $\omega$,
\begin{equation}
	\omega_{\protect\underline{k}}{}^i{}_j = - \omega_{\protect\underline{k}}{}^j{}_i \,.
\end{equation}
Here the higher-order compatibility with $G^i{}_j=0$ is easily achieved for the case of flat (zero-curvature)
gravitational fields while the general case demands some $G^i{}_j\neq0$.

It is not difficult to see that the condition (\ref{SpinConnectionFormulation}), after applying the frame postulate
\begin{equation}
	\partial_{\protect\underline{k}} e_{\protect\underline{l}}{}^i - \Gamma^{\protect\underline{i}}_{\protect\underline{kl}} e_{\protect\underline{i}}{}^i + \omega_{\protect\underline{k}}{}^i{}_j e_{\protect\underline{l}}{}^j = 0 \,,
\end{equation}
 can be equivalently rewritten, leaving out the derivatives of the vielbein, i.e only in terms of  $V^{\protect\underline{i}}$ and the Christoffel symbols.
\begin{align}
	& \partial_{\protect\underline{k}} V^{i} = (\partial_{\protect\underline{k}} e_{\protect\underline{i}}{}^i) V^{\protect\underline{i}} + e_{\protect\underline{i}}{}^i \partial_{\protect\underline{k}} V^{\protect\underline{i}}  = - \omega_{\protect\underline{k}}{}^i{}_j  e_{\protect\underline{i}}{}^j V^{\protect\underline{i}} + \eta_{\protect\underline{k}}{}^j G^i{}_j \implies \, \partial_{\protect\underline{k}} V^{\protect\underline{i}} = \Gamma^{\protect\underline{i}}_{\protect\underline{kl}} V^{\protect\underline{l}} + \eta_{\protect\underline{k}}{}^j G^{\protect\underline{i}}{}_j \,. \label{VdVGamma}
\end{align}
From here it is possible to use the metric formalism in the equations for $V^{\protect\underline{i}}$

\begin{equation}
	\d V^{\protect\underline{i}} = \kappa V^{\protect\underline{k}} \partial_{\protect\underline{k}} V^{\protect\underline{i}} + \eta^k \partial_{k} V^{\protect\underline{i}} = \kappa \Gamma^{\protect\underline{i}}_{\protect\underline{kl}}  V^{\protect\underline{k}} V^{\protect\underline{l}} +  \eta^j G^{\protect\underline{i}}{}_j \,,
\end{equation}
where
\begin{equation}
	\partial_k = e^{\protect\underline{k}}{}_k \partial_{\protect\underline{k}} \,.
\end{equation}
After the vielbein reduction on $V$: $P(e^i) = \kappa V^i$, one gets the expected geodesic equation.

The example above of the link between spin-connection and Christoffel symbols realizes the transition between worldsheet and fiber indices. This highlights the difference between the usual dynamics formulated in terms of $x^{\protect\underline{i}}, \dot x^{\protect\underline{k}}$ and our unfolded system formulated in terms of  $q^i, V^i$.

The relation between the two formalisms can be uplifted to the action level. As  noted in Section \ref{FreeParticle},  after an appropriate reduction of the vielbein, $Dq^i$ becomes $\frac{d q^i}{d \tau}$ ((\ref{Reductionq}), (\ref{Reductionv})), where $\tau$ is the natural evolution parameter. Using  that $q^i(x)$ are embedding functions for the coordinates $x^{\protect\underline{i}}$, let us write an action, quadratic in  $\frac{d q^i}{d \tau}$   using the metric $g_{ij}$ in the target space (not necessarily corresponding to Minkowski's space) and adding the gauge parameters $\alpha$ for reparametrization invariance by $\alpha' d\tau' = \alpha d\tau $,
\begin{align}
	& S = \frac{1}{2}\int d\tau \alpha \Big(\frac{1}{\alpha^2} g_{ij} \frac{d q^{i}}{d \tau} \frac{d q^{j}}{d \tau} + m^2\Big)
	\,.
\end{align}
In terms of $x^{\protect\underline{i}}$, we obtain the regular action
\begin{equation}
	S = \frac{1}{2} \int d\tau \alpha \Big (\frac{1}{\alpha^2} g_{ij} \frac{d q^{i}}{d x^{\protect\underline{k}}}  \frac{d q^{j}}{d x^{\protect\underline{l}}}\frac{d x^{\protect\underline{k}}}{d \tau}   \frac{d x^{\protect\underline{l}}}{d \tau} + m^2\Big ) = \frac{1}{2} \int d\tau \alpha \Big (\frac{1}{\alpha^2} \tilde{g}_{\protect\underline{kl}} \frac{d x^{\protect\underline{k}}}{d \tau}   \frac{d x^{\protect\underline{l}}}{d \tau} + m^2\Big) \,.
\end{equation}
Here $\tilde{g}_{\protect\underline{kl}}$ is the induced metric  from the target space. As usual, Euler-Lagrange equations for $\alpha$ are algebraic
\begin{equation}
	\alpha = \frac{1}{m} \sqrt{\tilde{g}_{\protect\underline{kl}} \frac{d x^{\protect\underline{k}}}{d \tau}   \frac{d x^{\protect\underline{l}}}{d \tau}} \,,
\end{equation}
which  allows us  to substitute them back into the action to arrive at the conventional result
\begin{equation}
	S = m \int d\tau  \sqrt{\tilde{g}_{\protect\underline{kl}} \frac{d x^{\protect\underline{k}}}{d \tau}   \frac{d x^{\protect\underline{l}}}{d \tau} } \,.
\end{equation}

\subsection{Interaction with higher spins}
Note that  the condition (\ref{SpinConnectionFormulation})  allows a direct generalization onto higher-spin interactions via introducing an appropriate higher-spin connection $\omega^{n_1, ... n_{s-1}, m}=dx^{\protect\underline{k}}\omega_{\protect\underline{k}}{}^{n_1, ... n_{s-1}, m}$ \cite{Vasiliev:1980},
\begin{align}
	& 	\d V^i = - \omega_{n_1, ... n_{s-1},}{}^{i} V^{n_1} ... V^{n_{s-1}} + \eta^j G^i{}_j \,.
\end{align}
In this case, the compatibility with the constraint (\ref{absV}) demands $\omega^{(n_1, ... n_{s-1}, m)} = 0$,
which means that, in agreement with the general higher-spin theory \cite{Vasiliev:1980}, $\omega^{n_1, ... n_{s-1}, m}$ is described by the Young diagram
\begin{ytableau}
	n_1 & ... & \scriptstyle n_{s-1} \cr m \cr
\end{ytableau}\,.

Note that the force associated with the field of an arbitrary spin can also be written in terms of generalized Christoffel symbols \cite{deWit:1979sib} (see also \cite{Segal:2000ke,Bars:2001um}).

\section{Conclusion} \label{Conclusions}

In this paper, we suggest an approach to the description of a relativistic classical  point particle
as a field on which
relativistic symmetries act geometrically. This is achieved by rewriting equations in the unfolded formalism
that supports manifest invariance under diffeomorphisms and the Lorentz group.
The point particle is represented as a field obeying unfolded equations.  A mechanism of projectors
 specifying  the evolution parameter in a covariant way is introduced.

The proposed approach can be useful for  different types of theories including the double field theory, \cite{Siegel:1993th, Siegel:1993xq} where, as we hope, it can be used to provide an alternative way of enforcing the section constraint (see, {\it e.g.}, \cite{Hohm:2017wtr}.) More generally,
the description of lower-dimensional objects within a proper extension of the proposed approach is of great
interest, in particular, for
 the description of branes in superstring theory as well as string theory itself:
it would be interesting to reformulate the string theory as a 2d theory described from the start in
terms of fields in the target space.
It would also be interesting to analyze the relation of the suggested mechanism with the models with non-linearly realized symmetries \cite{Ivanov:1975zq,Ivanov:1978mx}.

It should be stressed that in the presence of extra dimensions the unfolded dynamics solely along the
parameters associated with the particle trajectory does not allow for compatible unfolded  equations
demanding an evolution along the transverse directions respecting appropriate  compatibility
conditions. This raises a number of questions for the future study such as,
for instance, whether any solutions of the compatibility conditions can be associated with an evolution
along a one-dimensional trajectory and, if not, what are the sufficient conditions for this to be true?
A related interesting problem is to
obtain the on-shell conditions from the variational principle along the lines of \cite{Tarusov:2021fdk}.
%We plan to explore these interesting topics in a future publication.

\section*{Acknowledgement}
We are grateful to Ruslan Metsaev for the correspondence.
This work was supported by the Russian Basic Research Foundation Grant No 20-02-00208.

%\newpage


\begin{thebibliography}{99}
	\parindent=0pt
	\parskip=0pt


%\cite{Vasiliev:1988xc}
\bibitem{Vasiliev:1988xc}
  M.~A.~Vasiliev,
  %``Equations of Motion of Interacting Massless Fields of All Spins as a Free Differential Algebra,''
  Phys.\ Lett.\ B {\bf 209} (1988) 491.
  %doi:10.1016/0370-2693(88)91179-3

  %\cite{Vasiliev:1988sa}
\bibitem{Vasiliev:1988sa}
  M.~A.~Vasiliev,
  %``Consistent Equations for Interacting Massless Fields of All Spins in the First Order in Curvatures,''
  Annals Phys.\  {\bf 190} (1989) 59.
  %doi:10.1016/0003-4916(89)90261-3

%\cite{Vasiliev:2001dc}
\bibitem{Vasiliev:2001dc}
M.~Vasiliev,
%``Relativity, causality, locality, quantization and duality in the S(p)(2M) invariant generalized space-time,''
%doi:10.1142/9789812777065_0044
[arXiv:hep-th/0111119 [hep-th]].

%\cite{Vasiliev:2014vwa}
\bibitem{Vasiliev:2014vwa}
M.~Vasiliev,
%``Higher-Spin Theory and Space-Time Metamorphoses,''
Lect. Notes Phys. \textbf{892} (2015), 227-264
%doi:10.1007/978-3-319-10070-8_9
[arXiv:1404.1948 [hep-th]].

%\cite{Vasiliev:2001zy}
\bibitem{Vasiliev:2001zy}
M.~Vasiliev,
%``Conformal higher spin symmetries of 4-d massless supermultiplets and osp(L,2M) invariant equations in generalized (super)space,''
Phys. Rev. D \textbf{66} (2002), 066006
%doi:10.1103/PhysRevD.66.066006
[arXiv:hep-th/0106149 [hep-th]].

\bibitem{F}C.~Fronsdal, \emph{``Massless Particles, Ortosymplectic
Symmetry and Another Type of Kaluza-Klein Theory"}, Preprint
UCLA/85/TEP/10, in Essays on Supersymmetry, Reidel, 1986
(Mathematical Physics Studies, v.8).

\bibitem{BLS}
I.~Bandos, J.~Lukierski and D.~Sorokin, \textit{Phys.Rev.} {\bf D61} (2000) 045002,
{\tt [hep-th/9904109]}.

%\cite{Bandos:2005mb}
\bibitem{Bandos:2005mb}
I.~Bandos, X.~Bekaert, J.~de Azcarraga, D.~Sorokin and M.~Tsulaia,
%``Dynamics of higher spin fields and tensorial space,''
JHEP \textbf{05} (2005), 031
%doi:10.1088/1126-6708/2005/05/031
[arXiv:hep-th/0501113 [hep-th]].

\bibitem{Vinogradov}
Vinogradov, A.M. Geometry of nonlinear differential equations. J Math Sci 17, 1624–1649 (1981). https://doi.org/10.1007/BF01084594 .



%\cite{Batalin:1977pb}
\bibitem{Batalin:1977pb}
I.~A.~Batalin and G.~A.~Vilkovisky,
%``Relativistic S Matrix of Dynamical Systems with Boson and Fermion Constraints,''
Phys. Lett. B \textbf{69} (1977), 309-312.
%doi:10.1016/0370-2693(77)90553-6

%\cite{Barnich:2004cr}
\bibitem{Barnich:2004cr}
G.~Barnich, M.~Grigoriev, A.~Semikhatov and I.~Tipunin,
%``Parent field theory and unfolding in BRST first-quantized terms,''
Commun. Math. Phys. \textbf{260} (2005), 147-181
%doi:10.1007/s00220-005-1408-4
[arXiv:hep-th/0406192 [hep-th]].

%\cite{Barnich:2010sw}
\bibitem{Barnich:2010sw}
G.~Barnich and M.~Grigoriev,
%``First order parent formulation for generic gauge field theories,''
JHEP \textbf{01} (2011), 122
%doi:10.1007/JHEP01(2011)122
[arXiv:1009.0190 [hep-th]].

%\cite{Grigoriev:2012xg}
\bibitem{Grigoriev:2012xg}
M.~Grigoriev,
%``Parent formulations, frame-like Lagrangians, and generalized auxiliary fields,''
JHEP \textbf{12} (2012), 048
%doi:10.1007/JHEP12(2012)048
[arXiv:1204.1793 [hep-th]].

%\cite{Boulanger:2008up}
\bibitem{Boulanger:2008up}
N.~Boulanger, C.~Iazeolla and P.~Sundell,
%``Unfolding Mixed-Symmetry Fields in AdS and the BMV Conjecture: I. General Formalism,''
JHEP \textbf{07} (2009), 013
%doi:10.1088/1126-6708/2009/07/013
[arXiv:0812.3615 [hep-th]].

%\cite{Bekaert:2005vh}
\bibitem{Bekaert:2005vh}
  X.~Bekaert, S.~Cnockaert, C.~Iazeolla and M.~A.~Vasiliev,
  %``Nonlinear higher spin theories in various dimensions,''
  hep-th/0503128.

%\cite{Vasiliev:2005zu}
\bibitem{Vasiliev:2005zu}
M.~A.~Vasiliev,
%``Actions, charges and off-shell fields in the unfolded dynamics approach,''
Int.\ J.\ Geom.\ Meth.\ Mod.\ Phys.\  {\bf 3} (2006) 37
%doi:10.1142/S0219887806001016
[hep-th/0504090].

%\cite{Lada:1992wc}
\bibitem{Lada:1992wc}
T.~Lada and J.~Stasheff,
%``Introduction to SH Lie algebras for physicists,''
Int. J. Theor. Phys. \textbf{32} (1993), 1087-1104
%doi:10.1007/BF00671791
[arXiv:hep-th/9209099 [hep-th]].

%\cite{Shaynkman:2000ts}
\bibitem{Shaynkman:2000ts}
O.~Shaynkman and M.~A.~Vasiliev,
%``Scalar field in any dimension from the higher spin gauge theory perspective,''
Theor. Math. Phys. \textbf{123} (2000), 683-700
%doi:10.1007/BF02551402
[arXiv:hep-th/0003123 [hep-th]].


%\cite{Misuna:2022zjr}
\bibitem{Misuna:2022zjr}
N.~G.~Misuna,
%``On Unfolded Approach To Off-Shell Supersymmetric Models,''
[arXiv:2201.01674 [hep-th]].

\bibitem{Vasiliev:1980}
M. A. Vasiliev,
Yad. Fiz. 32 (1980) 855 [Sov. J. Nucl. Phys. 32 (1980) 439].

%\cite{deWit:1979sib}
\bibitem{deWit:1979sib}
B.~de Wit and D.~Z.~Freedman,
%``Systematics of Higher Spin Gauge Fields,''
Phys. Rev. D \textbf{21}, 358 (1980).

%\cite{Segal:2000ke}
\bibitem{Segal:2000ke}
A.~Y.~Segal,
%``Point particle in general background fields and generalized equivalence principle,''
[arXiv:hep-th/0008105 [hep-th]].

%\cite{Bars:2001um}
\bibitem{Bars:2001um}
I.~Bars and C.~Deliduman,
%``High spin gauge fields and two time physics,''
Phys. Rev. D \textbf{64} (2001), 045004
%doi:10.1103/PhysRevD.64.045004
[arXiv:hep-th/0103042 [hep-th]].

%\cite{Siegel:1993th}
\bibitem{Siegel:1993th}
W.~Siegel,
%``Superspace duality in low-energy superstrings,''
Phys. Rev. D \textbf{48}, 2826-2837 (1993).



%\cite{Siegel:1993xq}
\bibitem{Siegel:1993xq}
W.~Siegel,
%``Two vierbein formalism for string inspired axionic gravity,''
Phys. Rev. D \textbf{47}, 5453-5459 (1993).

%\cite{Hohm:2017wtr}
\bibitem{Hohm:2017wtr}
O.~Hohm, E.~T.~Musaev and H.~Samtleben,
%``O($d+1, d+1$) enhanced double field theory,''
JHEP \textbf{10} (2017), 086
%doi:10.1007/JHEP10(2017)086
[arXiv:1707.06693 [hep-th]].


%\cite{Ivanov:1975zq}
\bibitem{Ivanov:1975zq}
E.~Ivanov and V.~Ogievetsky,
%``The Inverse Higgs Phenomenon in Nonlinear Realizations,''
Teor. Mat. Fiz. \textbf{25} (1975), 164-177.

%\cite{Ivanov:1978mx}
\bibitem{Ivanov:1978mx}
E.~Ivanov and A.~Kapustnikov,
%``General Relationship Between Linear and Nonlinear Realizations of Supersymmetry,''
J. Phys. A \textbf{11} (1978), 2375-2384.

%\cite{Tarusov:2021fdk}
\bibitem{Tarusov:2021fdk}
A.~A.~Tarusov and M.~A.~Vasiliev,
%``On the variational principle in the unfolded dynamics,''
Phys. Lett. B \textbf{825} (2022), 136882
%doi:10.1016/j.physletb.2022.136882
[arXiv:2111.12691 [hep-th]].



\end{thebibliography}
\end{document}